\documentclass[twocolumn,floatfix, showpacs]{revtex4}

\usepackage[final]{epsfig}
\usepackage{amsmath}
\usepackage{parskip}

\begin{document}

\title{Mach cones in an evolving medium}

\author{Thorsten Renk, J\"{o}rg  Ruppert}

\pacs{25.75.-q}
\preprint{DUKE-TH-05-279}

\affiliation{Department of Physics, Duke University, PO Box 90305,  Durham, NC 27708 , USA}

\begin{abstract}
The energy and momentum lost by a hard parton propagating through hot and dense matter has to be redistributed in the nuclear medium. Apart from heating the medium, there is the possibility that collective modes are excited. 
We outline a formalism that can be used to track the propagation of such a mode
through the evolving medium if its dispersion relation is known. Under the assumption that a sound wave is created, 
we track the jet energy loss as a function of spacetime and follow the resulting mach cone throughout the fireball evolution.
We compare with the angular correlation pattern of hard hadrons as obtained by the PHENIX collaboration
and find good agreement with the data provided that a substantial fraction of jet energy ($\sim$ 90\%) is deposited
into a propagating mode and that the hot matter can be characterized by an EOS with a soft point (not necessarily a mixed phase).

\end{abstract}

\maketitle

\section{Introduction}
\label{sec_introduction}

Announcements have recently been made by the four detector collaborations at RHIC \cite{RHIC-QGP} that a new state of matter, distinct from ordinary
hadronic matter has been created in ultrarelativistic heavy-ion collisions (URHIC). A new and exciting challenge for both
experiment and theory is now to study its properties. The energy loss of hard partons created in the first moments
of the collision has long been regarded a promising tool for this purpose \cite{Jet1,Jet2,Jet3,Jet4,Jet5,Jet6}. However,
before this tool can be fully exploited, the influence of the medium evolution on the energy loss has to be understood
in a more quantitative way (cf. \cite{Jet-Flow}).

Recently, measurements of two-particle correlations involving one hard trigger particle have shown a surprising
splitting of the away side peak for all centralities but peripheral collisions, qualitatively very different from
a broadened away side peak observed in p-p or d-Au collisions \cite{PHENIX-2pc}. Interpretations in terms of energy
lost to propagating colourless \cite{Shuryak, Stoecker} and coloured \cite{Wake} sound modes have been suggested 
for this phenomenon. In \cite{Chaudhuri}, a Mach cone structure has been demonstrated to be visible in the angular pattern
of  photon emission if a source term for the medium energy is inserted in a hydro code, although this 
is at odds with the results of \cite{Shuryak} where this form of the source term does not yield a propagating shockwave.
In addition, we note that energy densities more than 20 times above the value of the surrounding medium in the vicinity of
1 fm around the jet as shown in Fig.~1 of \cite{Chaudhuri} do not seem to reflect the situation realized at RHIC, a 
possible reason for the discrepancy in the finding between \cite{Shuryak} and \cite{Chaudhuri}.

In the following, we investigate if such a mechanism can in agreement with the measured hadronic data provided 
that a realistic
evolution model of the hot and dense matter is used.

\section{Shockwave excitation, propagation and freeze-out}
Throughout this manuscript we focus on central collisions only.
Since we are interested in the deposition of lost jet energy
into the medium, our first task is to determine the spacetime pattern of energy loss.

For the description of the evolution, we base our investigation on
the fireball model outlined in \cite{RenkSpectraHBT} and \cite{Synopsis}. 
The model is tuned to describe transverse mass spectra of
hadrons, HBT correlation radii and hadronic $dN/dy$ distributions.
The model has successfully predicted the photon yield \cite{Photons_RHIC1, Photons_RHIC2, VNI-LPM} at RHIC
and gives a fair description of $R_{AA}$ in the high $p_T$ region without requiring unrealistically
high gluon multiplicities or transport coefficients far from the perturbative estimate \cite{Jet-Flow}. Therefore the model can be expected to give  a fair representation of the relevant physics in both 
early and late stages of the evolution.

\begin{figure}[!htb]
\vspace{0.5cm}
\epsfig{file=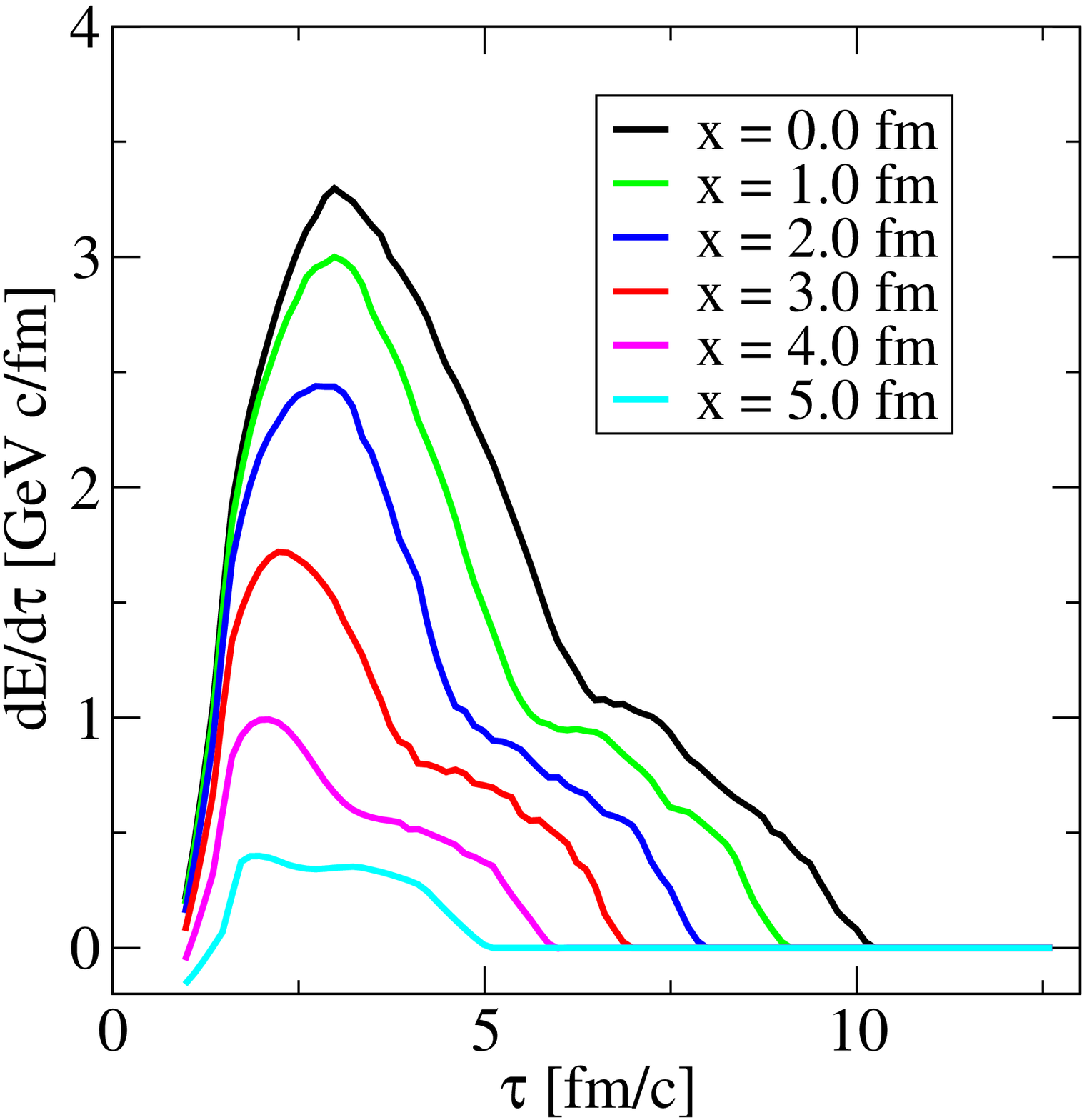, width=4.5cm}\epsfig{file=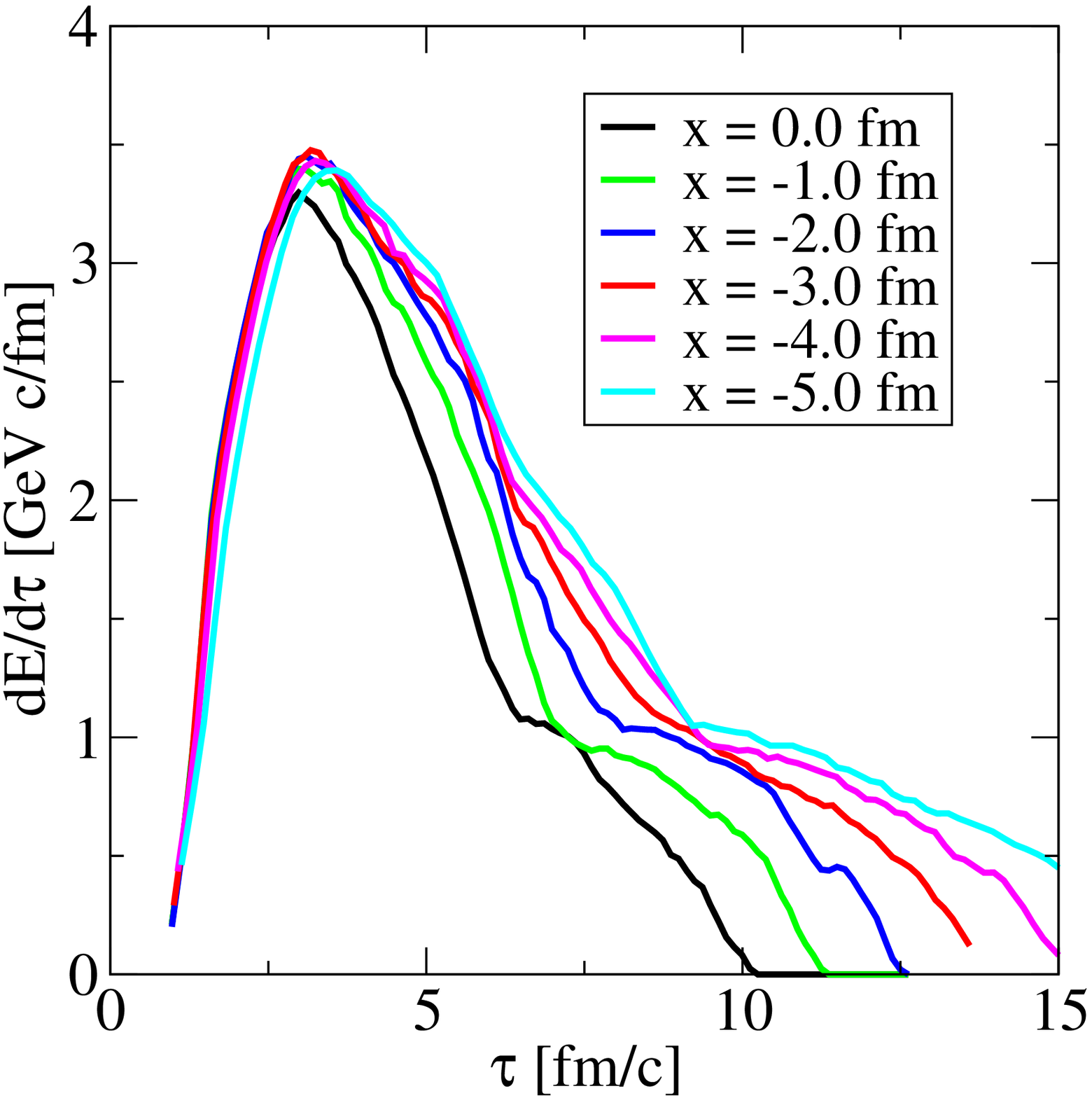, width=4.5cm}
\caption{\label{F-1}Energy deposition from hard quark at $\eta=0, y=0$  propagating into
positive $x$ as function of proper time in the CM frame $\tau$ for different initial positions $x$. The shape of the
energy loss pattern for gluons is qualitatively similar but has a larger absolute normalization.}
\end{figure}

For the energy loss calculation we use results of
\cite{QuenchingWeights} to obtain the probability $P(\Delta E)$ to lose the amount of energy $\Delta E$ from the two
key quantities, plasma frequency

\begin{equation}
\omega_c({\bf r_0}, \phi) = \int_0^\tau d \xi \xi \hat{q}(\xi)
\end{equation}
and averaged momentum transfer
\begin{equation}
(\hat{q}L) ({\bf r_0}, \phi) = \int_0^\tau d \xi \hat{q}(\xi)
\end{equation}

in a static equivalent scenario which are calculated along the path of the hard parton through the medium. 
Here we take $\hat{q}(\xi) = c \epsilon(\xi)^{3/4}$ with $\epsilon$ the local energy density of the medium 
as determined by the fireball evolution \cite{RenkSpectraHBT} and $c = 3$ as required
for the description of $R_{AA}$ \cite{Jet-Flow} and close to the perturbative $c\approx 2$ \cite{Baier}.
Since we are not interested in folding the result with a steeply falling spectrum but rather into the
energy deposited on average in a given volume element we focus on the average energy loss $\langle \Delta E \rangle =
\int_0^\infty P(\Delta E) \Delta E d\Delta E$ in the following.
For illustration of the resulting loss pattern,
we show the energy deposition for a hard parton at midrapidity ($\eta=0$), in the transverse $(x,y)$-plane at $y = 0$ and propagating into
positive $x$ direction as a function of initial position $x$ in Fig.~\ref{F-1}.

The result shows the initial rise of the differential energy loss as an approximately linear increase
reminiscent of the $L^2$ dependence of the energy loss in a static medium
and a rather sudden drop in $dE/dx$ once expansion significantly dilutes the medium. Once
transverse expansion starts becoming important, the high power of the volume growth with time wins out
over the advantage of having a longer decoherence time. It is the finite formation time of radiated quanta
which implies that lost energy does not appear in the medium immediately which causes the curves to rise from zero.

Our further investigation relies now on the assumption that a fraction $f$
of the energy lost to the medium excites a collective mode of the medium. Since our framework makes use of energy-momentum 
conservation laws, we do not have to address the important (though unsolved) question by what mechanism such a propagating mode is excited.  
We considered the case where colorless sound characterized by a dispersion relation $E = c_s^2 p$ is excited by that energy fraction $f$. The remaining energy fraction
$(1-f)$ in essence heats the medium and leads to some amount of collective drift along the jet axis to conserve longitudinal momentum.

\begin{figure}[!htb]
\vspace{0.5cm}
\epsfig{file=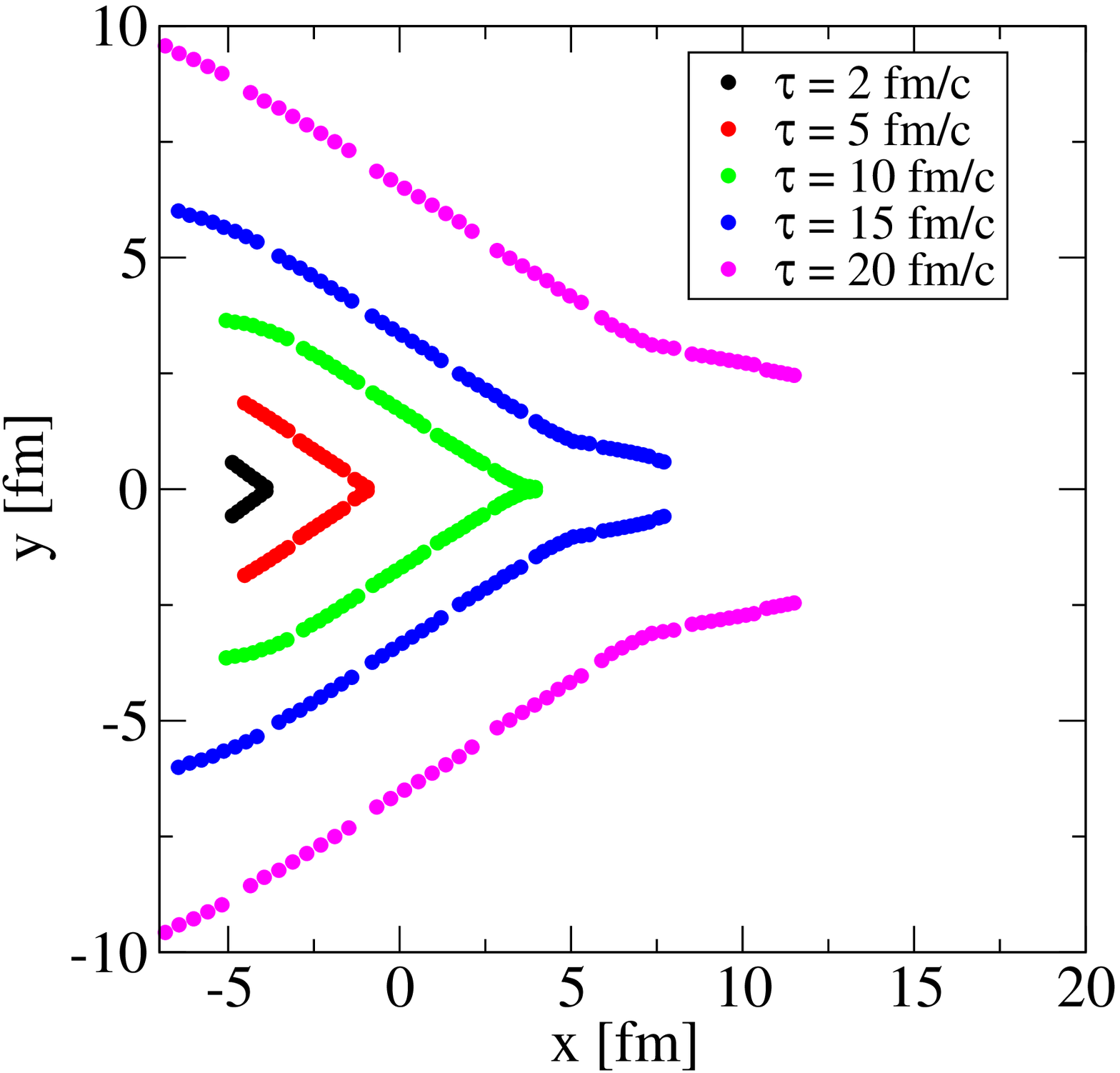, width=4.5cm}\epsfig{file=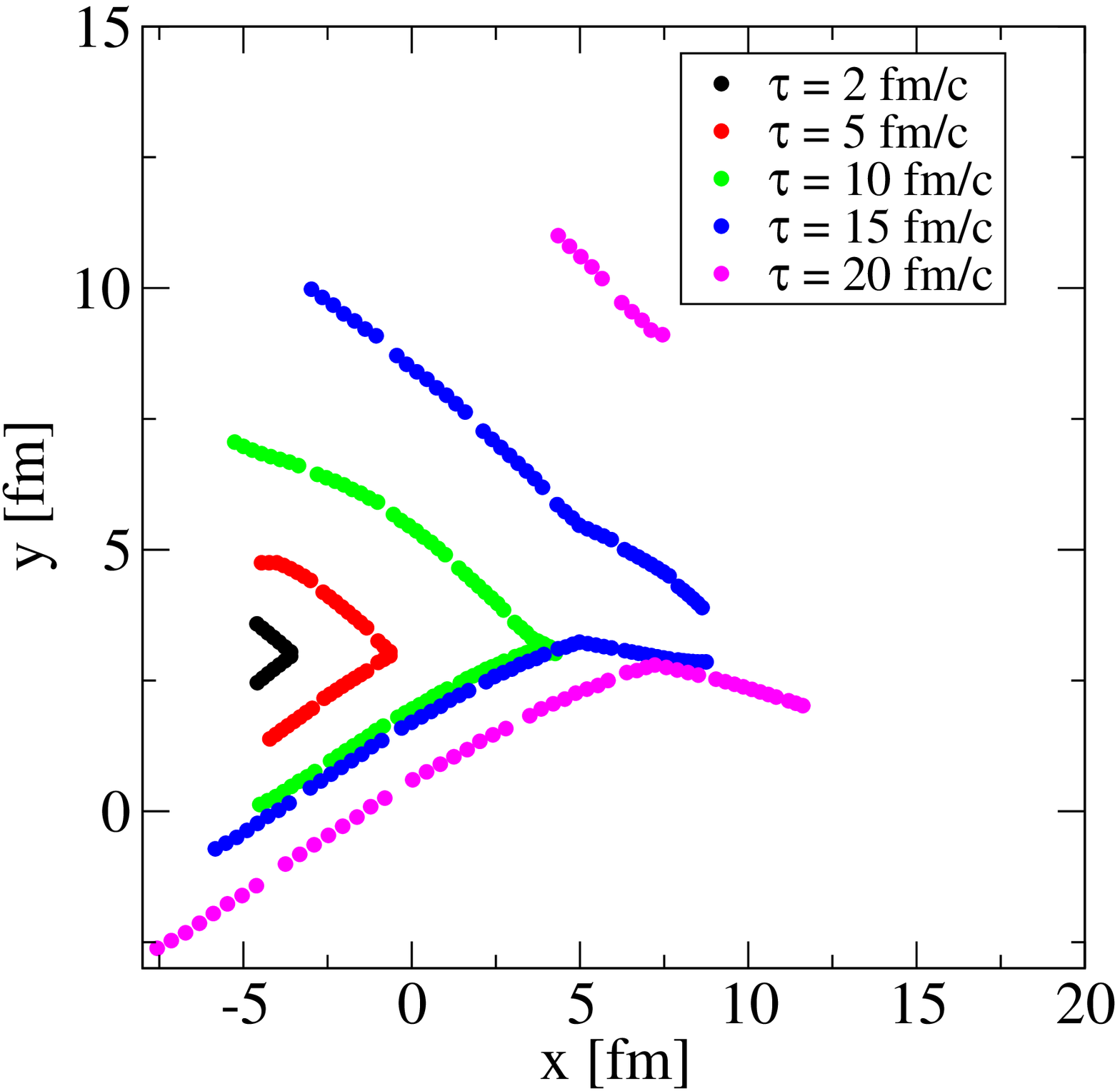, width=4.5cm}
\caption{\label{F-2}Mach cones excited by a jet travelling from $x=-5$ fm into positive $x$ direction
(with (0,0) the fireball center) for $y=0$ (left panel) and $y=3$ fm (right panel).}
\end{figure}

We calculate the speed of sound $c_s$ locally from a quasiparticle description of the equation of state as fitted to lattice results \cite{STW} as $c_s = \partial p(T)/ \partial \epsilon(T)$.
This EOS shows a significant reduction of $c_s$  as one approaches the phase transition but does not lead to a mixed phase.
The dispersion relation along with the energy and momentum deposition determines the initial angle of propagation
of the shock front with the jet axis (the 'Mach angle') as $\phi = \arccos c_s$.
We discretize the time into intervals $\Delta \tau$, calculate the energy deposited in that time interval
$E(\tau)$ and propagate the part of the shockfront remaining in the midrapidity
slice (i.e. in the detector acceptance).
Each piece of the front is propagated with the local speed of sound and the angle of propagation is
continously adjusted as 
\begin{equation}
\phi = \arccos \frac{\int_{\tau_E}^\tau c_s(\tau) d\tau }{(\tau - \tau_E)}
\end{equation}
where $c_s(\tau)$ is determined by the propagation history. Since sound propagates in the local
restframe, the shock front element is also carried away by the local flow velocity.
We illustrate this by showing the Mach cone in the transverse $\eta=0$ plane for a jet travelling
into positive $x$ direction, originating at $x=-6$ fm and either $y=0$ (i.e. going through the fireball
center) or $y=3$ fm in Fig.~\ref{F-2}. Both the effect of the soft point in the EOS narrowing the
cone at late times and the distortion of the cone in position space are clearly visible.
Note that the distortion by flow is sizeable in position space since flow velocities $v_T <0.7$ are
of the same order of magintude as the speed of sound $c_s \sim 0.3..0.5$. However, since the
effect of flow is already included in the standard calculation of $m_T$ spectra, the direction of
flow of excess momentum contained in the Mach cone in momentum space is hardly changed.

Once an element of the wavefront reaches the freeze-out condition $T = T_F$, a hydrodynamical
mode cannot propagate further. We assume that the energy contained in the shockwave
is not used to produce hadrons but rather is converted into kinetic energy. In the local
restframe, we then have a matching condition for the dispersion relations
\begin{equation}
E = c_s p \quad \text{and} \quad E = \sqrt{M^2 + p^2} - M
\end{equation} 
where $M = V \left(p(T_F) + \epsilon(T_F)\right)$ is the 'mass' of a volume element at freeze-out temperature. 

Once we have calculated the additional boost a volume element receives from the shockwave using the
matching conditions, we use the Cooper-Frye formula
\begin{equation}
E \frac{d^3N}{d^3p} =\frac{g}{(2\pi)^3} \int d\sigma_\mu p^\mu
\exp\left[\frac{p^\mu u_\mu - \mu_i}{T_f}\right] = d^4 x S(x,p)
\end{equation}
with $u^\mu = u^\mu_{flow} + u^\mu_{shock}$ to convert the fluid element into a hadronic distribution.
The resulting momentum spectrum is thus a thermal two component spectrum resulting from an integrations involving
volume not part of the shockwave and volume receiving an additional boost from the shockwave.

From those considerations one can also calculate the typcial number particles getting an additional boost from the shock front. 
An upper bound for $c_s$ at $T_F$ would be the speed of sound in a hadronic resonance gas $c_s \approx \sqrt{0.2}$ 
\cite{HadSound} with a freeze-out temperature of $\approx 110~{\rm MeV}$ \cite{RenkSpectraHBT}. From those considerations one 
finds for an energy in the collective mode of about $5~{\rm GeV}$ a number of $\approx 20$ pions
which are part of the shockwave. Assuming  e.g. $c_s=0.3$ one finds $\approx 50$ pions. 
The mach cone would also contain a smaller number of additional kaons and nucleons. 
A $10$ GeV jet can therefore potentially involve $\mathcal{O}(100)$ particles. 

\section{Sampling the medium evolution}

In the following we compare to the hard two particle correlation data. For this purpose we simulate the 
PHENIX trigger conditions as closely as possible using a Monte-Carlo approach. 
We start by generating vertices with a distribution weighted by the nuclear overlap
\begin{equation}
T_{AA}({\bf b}) = \int dz \rho^2({\bf b},z).
\end{equation}

We then determine the jet momentum and parton type by randomly sampling partonic transverse momentum spectra generated
by the VNI/BMS parton cascade as described in \cite{VNI-LPM}. Calculating the energy
loss of the near side parton, we decide if the experimental trigger condition is fulfilled.
Since the experiment triggers on a hard hadron in the transition between the recombination and fragmentation regime, 
the model at this point cannot implement the trigger condition exactly. Instead, we require the trigger
condition to be fulfilled by the parton and have checked that the model results do not
change significantly when the near side trigger threshold is increased by 2 GeV.
We note that this procedure places the vertices fulfilling the trigger condition close to the surface 
of the produced matter, i.e. in our model the
medium is rather opaque, in agreement with the conclusions of \cite{Fragility}.

Once the vertex and momentum of a near side jet has passed the trigger condition, we 
determine the direction of the far side parton in the transverse plane. In order to
take into account intrinsic $k_T$, we do not propagate the away side directly opposite to the near side jet but allow for a random angle. We have verified that this distribution, folded with the width of the near
side peak reproduces the width of the far side peak in the case of d-Au and 60-90\% peripheral Au-Au collisions.

With vertex, energy and direction of the away side jet fixed, we calculate $dE/d\tau$ of the outgoing parton.
We stop the calculation when a significant fraction of the energy is lost to the medium.
We generate and propagate a shockwave by the formalism outlined in the previous section and fold the
resulting hadron spectra with the experimental acceptance cuts. 
For the remnants of the away side jet, thermal physics and local boosts
determine the spectral excess, therefore we have no ambiguity assoicated with the choice of 
recombination/fragmentation.

In each event we assume that a fraction $(1-f)$ of the energy lost from the hard parton heats the medium.
We assume that due to momentum conservation this will lead to additional flow into the direction of the
original away side parton. Likewise, we account for the possibility of a punchthrough if the initial 
vertex is very peripheral and  both near and away side parton propagate near-tangential to the 
surface (however we note that these events are extremely rare and play almost no role in the calculation).

\section{Results}

\begin{figure}[!htb]
\epsfig{file=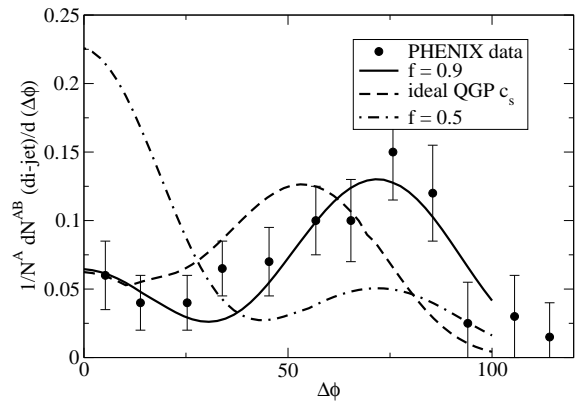, width=7.5cm}
\caption{\label{F-3}Away side correlation structure for different scenarios.}
\end{figure}

We present the resulting 2-particle correlation structure on the away side in Fig.~\ref{F-3}. Since the 
near side associated particle distribution is not calculable in the framework outlined above we will
only focus on the away side region and make no attempt to describe near side correlations, hence our
results are normalized to the integral of the away side correlation structure. 
We have chosen 0 degrees as the direction opposite to the near side jet.

We present three different scenarios: Our best fit of the only model parameter $f$ to the data (leading to
$f \approx 0.9$)
a scenario where $f$ is set to 0.5 and one in which we do not determine $c_s$ by the EOS but set it to
the ideal QGP value $c_s \approx 1/\sqrt{3}$.

One can read off two conclusions immediately from the comparison with the data: First, a large fraction (about 90\%)
of the lost energy has to excite a shockwave in order to reproduce the data. Since most of the 'Mach ring' in
momentum space is outside the acceptance but any flow or punchthrough is predominantly inside, even a
moderate fraction of jet energy heating the plasma would lead to a strong signal at zero degrees. In order
to find a dip, the mechanism transporting energy away from this region has to be very efficient.

Second, the angular position of the peak is capable of measuring the speed of sound to some degree. We observe
that on average it has to be significantly smaller than the ideal QGP value and that the data are
in agreement with the situation as seen on the lattice, i.e. a soft point but no mixed phase. However, it
would be premature to claim that this is the only possibility to describe the data.

\section{Summary and outlook}

We have demonstrated that the idea of shockwaves being excited by the energy lost from a hard parton
into the medium is able to account for the observed splitting of the away side peak if a realistic model
for the medium evolution is used. If this is the underlying mechanism, it has to dominate over
simple heating of the plasma, as this would produce a significant peak in forward direction.
In this scenario, the away side peaks essentially represent thermal physics, and we expect the momentum
distribution at the peak between 1 and 2 GeV to be dominated by boosted thermal spectra and not by jet fragmentation
or recombination physics.

We emphasize that the framework we have used is quite general and relies mainly on 
energy momentum conservation and the assumption of a deposition of a considerable fraction 
of the energy in a collective mode.  We used a specific dispersion relation (colorless sound) 
to demonstrate that the two particle correlations with hard trigger have some sensitivity to $c_s$ 
in hot and dense matter.
We see an average $c_s$ far from the ideal QGP value and well compatible with a crossover transition with a soft point.

Recently, measurements of 3-particle correlations have been presented by both the PHENIX \cite{PHENIX-3p}
and the STAR \cite{STAR-3p} collaboration. While we plan to address a calculation of 3-particle correlations
in the framework outlined here in a future publication, we'd like to remark at this point that since
the Mach cone is a collective phenomenon that can - as discussed above - involve  typically of the order of 20-50 particles
for the given trigger conditions.
While all of them are strongly correlated with the original away-side parton (and hence with the
observable near side jet), once this correlation is subtracted only the momentum balance (with respect to
the away-side parton momentum) $\sum p_L = p_{away}$ and $\sum {\bf p_T} = 0$ remains to account for residual
correlation inside the cone. With momentum distributed across a significant number of particles, there is no a priori reason
to expect strong correlations between two specific ones.

There are still many significant physics questions regarding  jet energy deposition into
the medium: 
Is there an appropiate theoretical framework to understand the jet energy deposition mechanism of a fraction $f$ of the lost energy into collective modes in detail?
Is sound the only collective mode compatible with the data? How can colorful and colorless sound collective modes and their interplay be consistently included? How sensitive is the correlation pattern
to details of the fireball evolution or the EOS? Can we successfully calculate the $p_T$ spectra of near and
far side simultaneously? At the moment, the case for the excitation and propagation of a sound mode through the 
hot and dense system appears to be rather strong and offers exciting opportunities to study the evolving system
from a new angle, and we plan to address those questions in forthcoming publications along with a 
systematic investigation of some of our model assumptions.

\begin{acknowledgments}
We would like to thank R.~Lacey, S.~A.~Bass and B.~M\"{u}ller for helpful discussions, comments and their
support during the preparation of this paper.

This work was supported by the DOE grant DE-FG02-96ER40945 and two Feodor
Lynen Fellowship of the Alexander von Humboldt Foundation.
\end{acknowledgments}

\end{document}